\documentclass[5p]{elsarticle}

\usepackage{lineno,hyperref}
\usepackage{amsmath}
\usepackage{amssymb}
\usepackage{xspace}
\modulolinenumbers[5]
\usepackage{graphicx}
\usepackage{bm}
\usepackage[justification=raggedright]{caption}
\usepackage{subcaption}
\usepackage{physics}
\usepackage{algorithm}
\usepackage{algorithmic}
\usepackage{comment}
\usepackage{xcolor} 
\usepackage{hyperref}
\usepackage{empheq}
\usepackage[british]{babel}
\usepackage{csquotes}
\usepackage{gensymb}
\usepackage{caption}
\usepackage{subcaption}

\journal{Nuclear Materials and Energy}

\begin{document}
	
\begin{frontmatter}
\title{Inferring the scrape-off layer heat flux width in a divertor with a low degree of axisymmetry}

\author[TE]{C Marsden}
\author[TE]{X Zhang}
\author[TE]{M Moscheni}
\author[ORNL]{T K Gray}
\author[TE]{E Vekshina}
\author[TE]{A Rengle}
\author[LF]{A Scarabosio}
\author[TE]{M Sertoli}
\author[TE]{M Romanelli}
\author[TE]{and the ST40 team}
\address[TE]{Tokamak Energy Ltd., 173 Brook Drive, Milton Park, Abingdon, UK}
\address[ORNL]{Fusion Energy Division, Oak Ridge National Laboratory, USA}
\address[LF]{LINKS Foundation, Via Pier Carlo Boggio 61, Torino, 10138, ITA}

\begin{abstract}
Plasma facing components (PFCs) in the next generation of tokamak devices will operate in challenging environments, with heat loads predicted to exceed 10 MWm$^{-2}$. The magnitude of these heat loads is set by the width of the channel, the ‘scrape-off layer’ (SOL), into which heat is exhausted, and can be characterised by an e-folding length scale for the decay of heat flux across the channel. It is expected this channel will narrow as tokamaks move towards reactor relevant conditions. Understanding the processes involved in setting the SOL heat flux width is imperative to be able to predict the heat loads PFCs must handle in future devices. Measurements of the SOL width are performed on the high-field spherical tokamak, ST40, using a newly commissioned infrared thermography system. With its high on-axis toroidal magnetic field ($\geq$1.5 T) ST40 is uniquely positioned to investigate the influence of toroidal field on the heat flux width in spherical tokamaks, whilst also extending measurements of the SOL width in spherical tokamaks to increased poloidal field ($\geq$0.3 T). Due to the divertor on ST40 having a low degree of axisymmetry, it is necessary for a set of radial measurements of the heat flux to be taken across the divertor, made possible using an automated toolchain that fully incorporates its 3D geometry. These radial profiles are combined with the magnetic topology of the plasma to infer the width of the SOL, with both single and double exponential profiles of heat flux observed. A reduction in the heat flux is observed toroidally across part of the divertor, with preliminary investigations indicating that partial shadowing occurs, resulting from the separation between magnetic field lines and trailing edges upstream of the observed region becoming comparable to the ion gyro-radius.

\end{abstract}

\end{frontmatter}

\section{Introduction}

The power exhausted in the next generation of tokamaks poses a challenge to the survivability of divertor plasma facing components (PFCs). Devices such as ITER are expected to face steady-state heat loads $\geq$10 MWm$^{-2}$\cite{pitts2019physics}, with the SPARC device facing heat loads  $\geq$100 MWm$^{-2}$ for $\sim$10 s\cite{kuang2020divertor}. For a robust design of the divertor to be realised, a load case for the heat flux onto the surface of these PFCs must be constructed. A cornerstone of such predictive modelling is an estimation of the width of the the heat exhaust channel, the \enquote{scrape-off layer} (SOL), into which power will be exhausted, before being transported down to the divertor. The width of the SOL will determine how concentrated the resultant heat loads are, and hence is responsible for setting the level of engineering challenge in the design of the divertor.\newline

As it presently stands, there is a lack of a complete first principles understanding of the myriad of physical processes at play that determine the SOL width. This poses a challenge for the design of the next generation of tokamaks, as the SOL width must be estimated, such that load cases can be created that will then drive the engineering design of the PFCs. It is therefore important that the SOL width of existing fusion experiments be measured, such that our theoretical understanding of the processes at play can continue to be developed. In the absence of such a first principles model, it has become desirable to make predictions of the SOL width using scaling laws derived from experimentally measured values from existing fusion experiments and the associated plasma and machine parameters. Inference of the SOL width is typically carried out by at least one of the following diagnostic techniques; infrared (IR) thermography\cite{vondracek2019divertor}, Langmuir probe measurement\cite{adamek2017electron} or Thomson scattering measurement\cite{hecko2023experimental}.\newline

\begin{figure*}[hbt!]
    \centering
    \begin{subfigure}[b]{0.47\linewidth}
        \centering
        \includegraphics[width=\textwidth,angle=180]{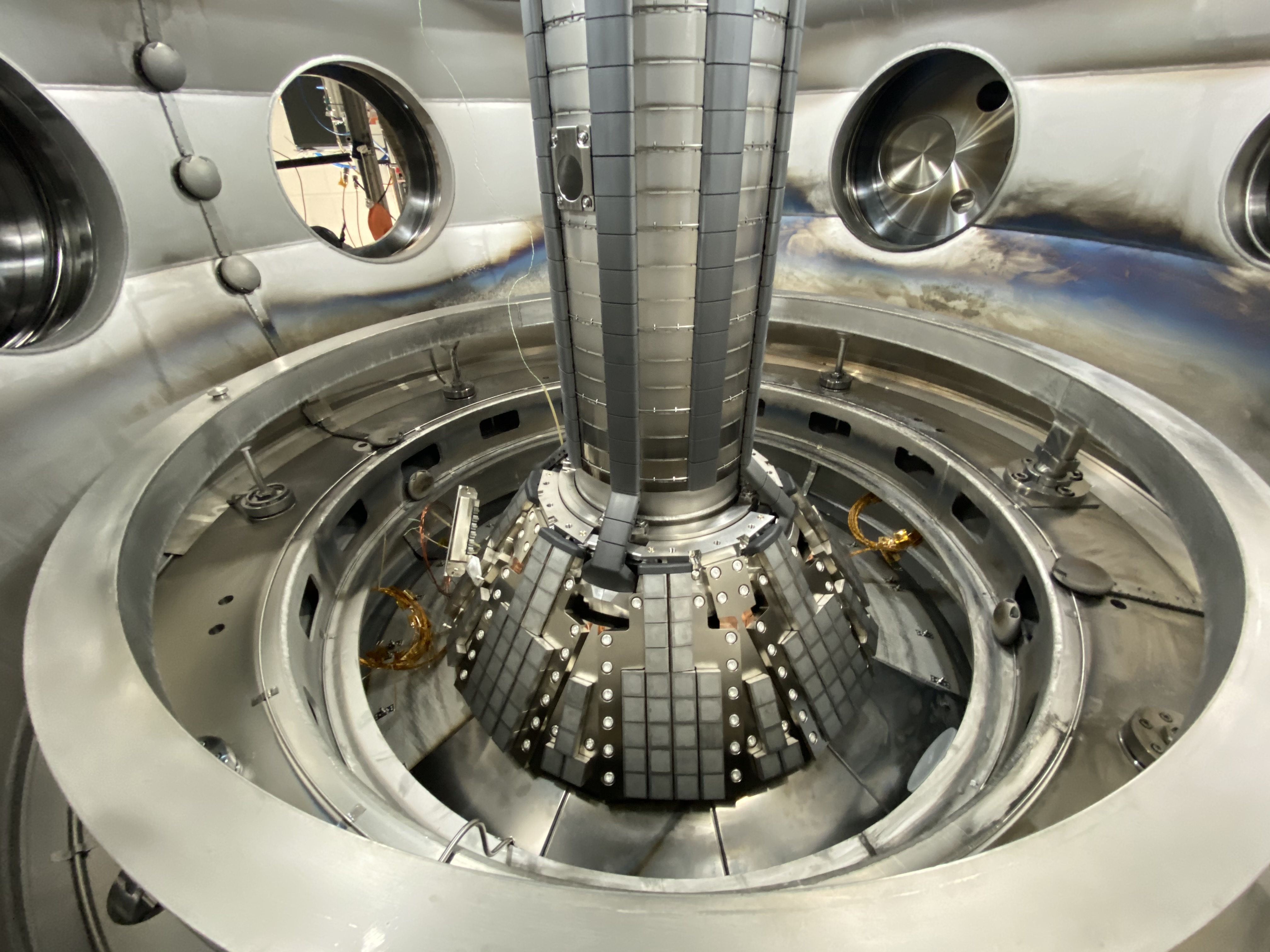}
        \caption{Upper divertor in-vessel view}
        \label{fig:ST40_and_CAD_1}
    \end{subfigure}
    \hfill
    \begin{subfigure}[b]{0.52\linewidth}
        \centering
        \includegraphics[width=\textwidth]{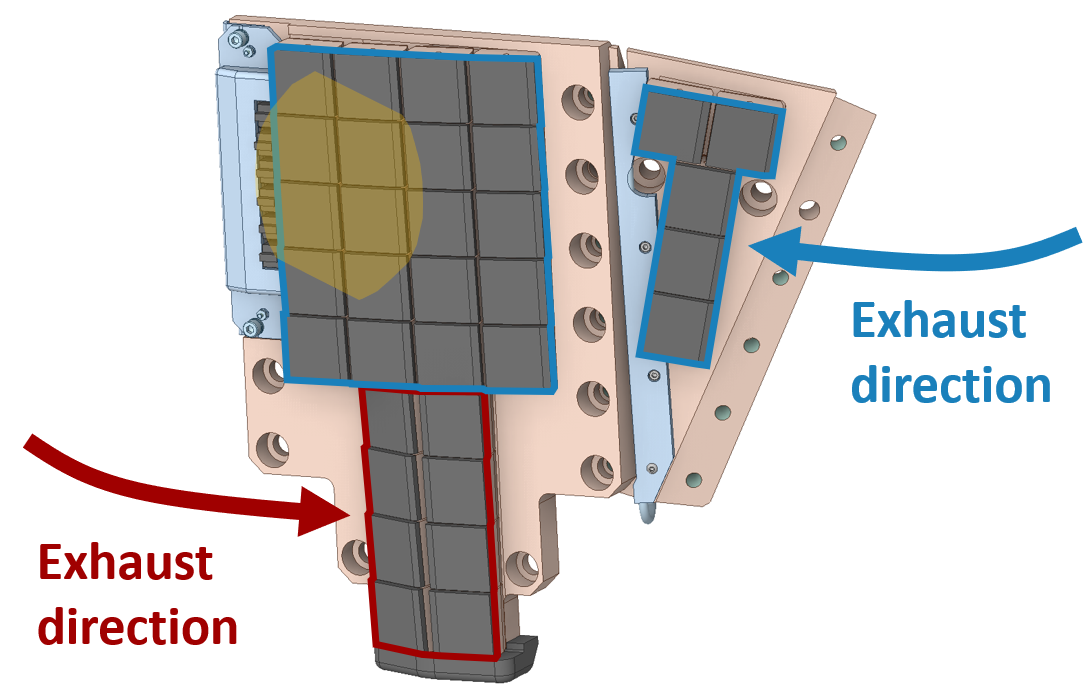}
        \caption{CAD model of the upper divertor}
        \label{fig:ST40_and_CAD_2}
    \end{subfigure}
        \caption{\ref{fig:ST40_and_CAD_1} In-vessel view of the upper divertor taken during installation work in July 2020. \ref{fig:ST40_and_CAD_2} Segment of the ST40 upper divertor. The outer divertor region on the low-field side is highlighted in blue, with tiles tilted in the opposite direction to those in the inner divertor region on the high-field side, highlighted in red. Arrows indicate the opposite toroidal directions from which the exhaust plasma strikes the tiles. Unique to this segment are a set of 6 Langmuir probes adjacent to the trailing edge of the LFS tiles, as well as an IR camera which views the region highlighted in orange.}
        \label{fig:ST40_and_CAD}
\end{figure*}

The IR thermography approach involves observing a region, in this case the surface of the divertor, with an infrared camera. The temperature of this surface is measured, before a thermal model of the PFCs is employed in which the heat equation can be solved for the observed temperature trace. This procedure will return the incident heat flux normal to the PFC surface, $q_{\perp}$. These measurements of $q_{\perp}$ can then be combined with measurements of the plasma's magnetic geometry and information about the shape of the PFCs in order to infer the SOL width.

\section{The ST40 divertor}
IR camera thermography is employed on ST40\cite{mcnamara2024st40}, a high-field spherical tokamak operated by Tokamak Energy Ltd. ST40 typically operates in a double-null diverted configuration, and is equipped with an up-down symmetric pair of divertors. Each divertor is divided into 8 segments, each of which consists of a pair of PFCs; one larger and one smaller set of molybdenum tiles brazed onto CuCrZr bases which are in turn bolted onto a copper carrier that is coated with nickel\cite{bamber2021st40}. On the larger carrier sit tiles for both the inner divertor located on the high-field (HFS), and the outer divertor located on the low-field side (LFS) of the machine, with the tiles tilted in opposite toroidal directions so as to cast magnetic shadows that protect the leading edges of adjacent tiles in a so-called \enquote*{fish-scaled} configuration. The leading edges of each tile are further protected by a poloidal chamfer that acts to reduce the near normal angle at which the exhaust plasma may otherwise strike the sides of the tiles at.\newline

ST40 is equipped with a 640$\times$512 px, 800 Hz divertor IR camera, calibrated in the 20-113.6 $\degree$C temperature range. Via an endoscope, this IR camera views a portion of the outer divertor target in the upper divertor region of the machine, as shown in figure \ref{fig:ST40_and_CAD}. The spatial resolution of the image of the divertor surface is $\sim$0.2 px/mm, enabling small-scale features in the tile surface temperature to be well resolved. It was found during the last campaign that temperatures recorded by the camera would saturate towards the end of the higher performance pulses, with it having since been re-calibrated for a maximum observable temperature of 200$\degree$C ready for the next campaign.

\section{Measuring the divertor surface heat flux}
\begin{figure}
    \centering
    \includegraphics[width = 0.95\linewidth]{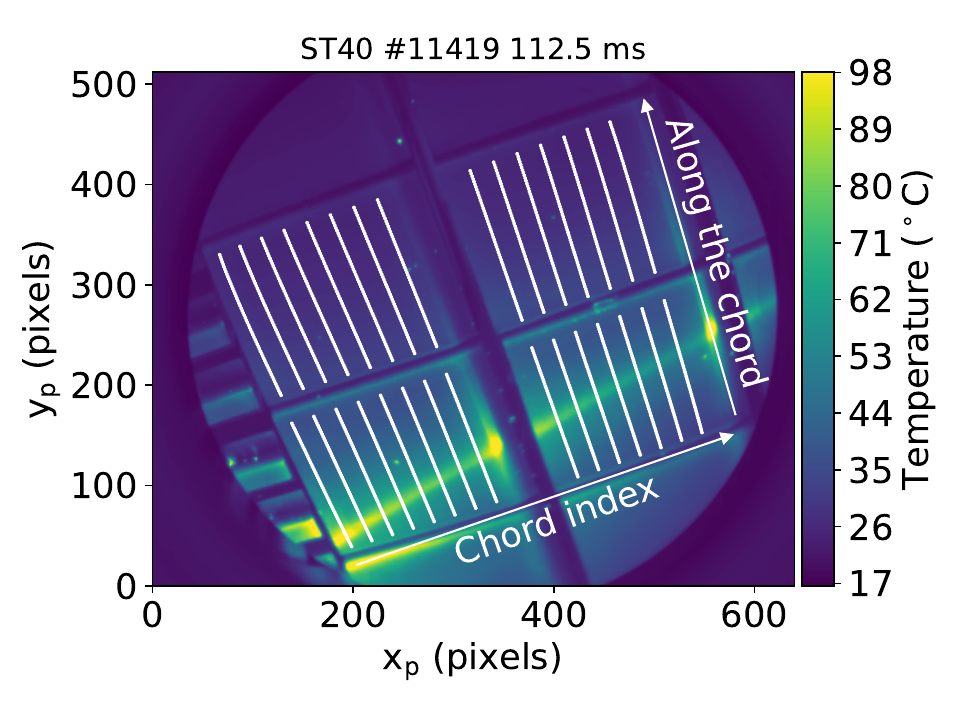}
    \caption{IR camera view of the ST40 upper divertor LFS region. Pixel colours correspond to the observed temperature, with a series of observation chords along the tile surfaces shown in white. 4 tiles are viewed in their entirety, referred to herein as a left-hand pair and right-hand pair of tiles. 60 chords are taken across the toroidal extent of the divertor, with 30 chords on each tile pair. A subset of these chords are shown for visual clarity.}
    \label{fig:ST40_divertor_chords}
\end{figure}

The divertor surface heat flux is calculated from the tile surface temperatures recorded by the IR camera. The first step in this procedure is to map the $(x_{p},y_{p})$ pixels of the IR camera image to the (X,Y,Z) Cartesian coordinates of the divertor. This is achieved using Calcam\cite{calcam}, wherein a set of distinctive points (tile and probe corners) on the divertor are manually associated with corresponding pixels on the camera image. With the mapping created, it is now possible to, for any (X,Y,Z) point on the divertor, interpolate the corresponding $(x_{p},y_{p})$ position in the IR camera image. Next, a series of \enquote*{observation chords} are drawn along the surfaces of each tile. These are uniformly spaced across the tiles, with a small offset from the edges set. Each chord spans a pair of tiles, allowing for a set of radial profiles of the divertor surface temperature to be taken. These observation chords can be seen in figure \ref{fig:ST40_divertor_chords}. With a 1D surface temperature profile along each chord taken, a 2D thermal model of the divertor tiles is employed for each chord. This thermal model, FAHF (the Functional Analysis of Heat Flux), was developed in-house at Tokamak Energy\cite{robinson2024fahf}. It solves the heat equation in 2D using an explicit finite differences scheme, with the first dimension being along the divertor surface (along the chord) and the second dimension down into the depth of the tile. By solving the heat equation, the surface heat flux along each chord is obtained.

\section{Data quality checks}
In order to infer the width of the SOL from divertor surface heat flux measurements, an IR analysis code - IRRITANT (the InfRaRed InvestigaTive ANalysis Toolchain) was developed. For each IR image taken during a pulse, the corresponding set of 1D divertor surface heat flux measurements along each observation chord are analysed. IRRITANT can perform this analysis on any set of 1D surface heat flux measurements; it is not required that FAHF specifically be used to produce these heat flux data, rather, any thermal model can be used that is capable of producing radial profiles of surface heat flux. The first part of this analysis is a data quality check of both the surface temperature and heat flux data on each chord, as well as a series of checks on the plasma equilibrium that has been reconstructed using the code EFIT\cite{lao1985reconstruction}.\newline

The surface temperature is checked for saturation, with chords found to contain saturated temperature data discarded. A check on the distance between the point of peak surface heat flux and the edges of the tiles is performed, so as to ensure that the true point of peak heat flux is not hidden down a tile gap that is not observable. This check is important as the region around the point of peak heat flux will contain data that is crucial for inferring the width of the SOL. Next, a check is performed on how much the heat flux decreases when moving away from its peak value. If the drop-off in heat flux on one side of the peak is small compared with how much the heat flux drops-off on the other side of the peak, then this indicates that the heat flux footprint is being truncated preferentially on one side - either in the common flux region (CFR) or in the private flux region (PFR). Different functional forms will later be fit to these data in order to extract the SOL width, with certain functions requiring that both the CFR and PFR regions are in view in order to fit the data.

A series of checks on the equilibrium plasma are also carried out. The plasma must be diverted at the time instance being analysed, with the separatrix through the upper divertor region's X-point in view of the region observed by the IR camera. The plasma current is checked to ensure it is in its flat-top phase (within 10$\%$ of the mean value during the flat-top), with the vertical position of the magnetic axis also examined in order to ensure that the plasma is not undergoing a vertical displacement event. Lastly, the state of the 2 neutral beams on ST40 are examined, in order to avoid analysing frames close to times in which the heating and current drive from the beams suddenly changes.

\section{Mapping surface to parallel heat flux}
With the appropriate data quality checks performed, the measurements of the divertor surface heat flux $q_{\perp}$ taken along each chord can be further processed. In order to measure the SOL width, the heat flux parallel to the magnetic field lines on the divertor target surface, $q_{\parallel,t}$, must be computed (here, the subscript \enquote{t} denotes the target position, i.e. the surface of the divertor). This is achieved by using the full 3D incidence angle, $\gamma$, of the magnetic field lines into the divertor, such that

\begin{equation}
    q_{\parallel,t} = \frac{q_{\perp}}{sin(\gamma)} = \frac{q_{\perp}}{\hat{b} \cdot \hat{n}}
\end{equation}

where $\hat{n}$ is the surface normal unit vector at a point on the divertor. $\hat{b}$ is the magnetic field line unit vector at this same point and is taken from the reconstructed equilibrium plasma. The divertor surface normal $\hat{n}$ could be calculated simply using the 3 corner points that define the flat tile surface on which a point sits, however, it was deemed desirable for IRRITANT to be extendable to any PFC surface topology, and so a more general approach is taken. This approach involves taking a high resolution mesh of the PFC surface, such that the normal vector over the surface can be computed. In order to mesh the PFCs, the open-source CAD programme FreeCAD\cite{FreeCAD} was used. FreeCAD allows the user to create a surface mesh of a supplied CAD model of the PFC of interest, and does this via the use of the meshing algorithm Mefisto\cite{Perronnet}. Mefisto creates a triangular mesh whose maximum edge length is specified by the user, which has the advantage of being able to maintain a consistently high resolution across the entire PFC surface. With the divertor surface normals obtained, the calculation of $\gamma$ is made possible, enabling $q_{\parallel,t}$ to be computed.

In order to infer the SOL width, these parallel heat flux data at the divertor surface are mapped to the LFS outboard magnetic midplane (OMP). Those points along an observation chord that are in the common flux region sit at the end of a magnetic flux tube that connects points on the divertor surface to points along the OMP. The heat flux parallel to the field line embedded at the centre of this flux tube is inversely proportional to the cross sectional area of the flux tube, which itself is inversely proportional to the strength of the magnetic field at a given point along the tube. This variation in magnetic field strength between the OMP and the divertor target is known as the total flux expansion, $F_{x}$, defined as

\begin{equation}
    F_{x} = \frac{B_{u}}{B_{t}}
\end{equation}

where the subscript \enquote{u} denotes the upstream point - in this case along the OMP. In order to create a mapping between points on the divertor and their corresponding upstream points, the poloidal magnetic flux, $\psi$, is sampled along the OMP, such that a mapping between $\psi$ and the corresponding R coordinate along the OMP is formed. From the reconstructed equilibrium plasma, $\psi$ at each point on the divertor surface is known, hence the corresponding upstream points along the OMP can be determined. With these upstream points, the parallel heat flux as mapped upstream, $q_{\parallel,u}$, can be computed

\begin{equation}
    q_{\parallel,u} = q_{\parallel,t} F_{x}
\end{equation}

It is important to reiterate that this mapped heat flux is not itself the actual parallel heat flux along the OMP, rather, this is the parallel heat flux at the divertor mapped to the OMP. The splitting of power between the different divertor legs, as well as radiative losses between the OMP and divertor target are not accounted for. Moreover, the effect of cross-field transport localised to the divertor region is still present in the mapped heat flux profile.

\section{Extracting the SOL width from profiles of parallel heat flux}

\begin{figure}
    \centering
    \includegraphics[width = 0.95\linewidth]{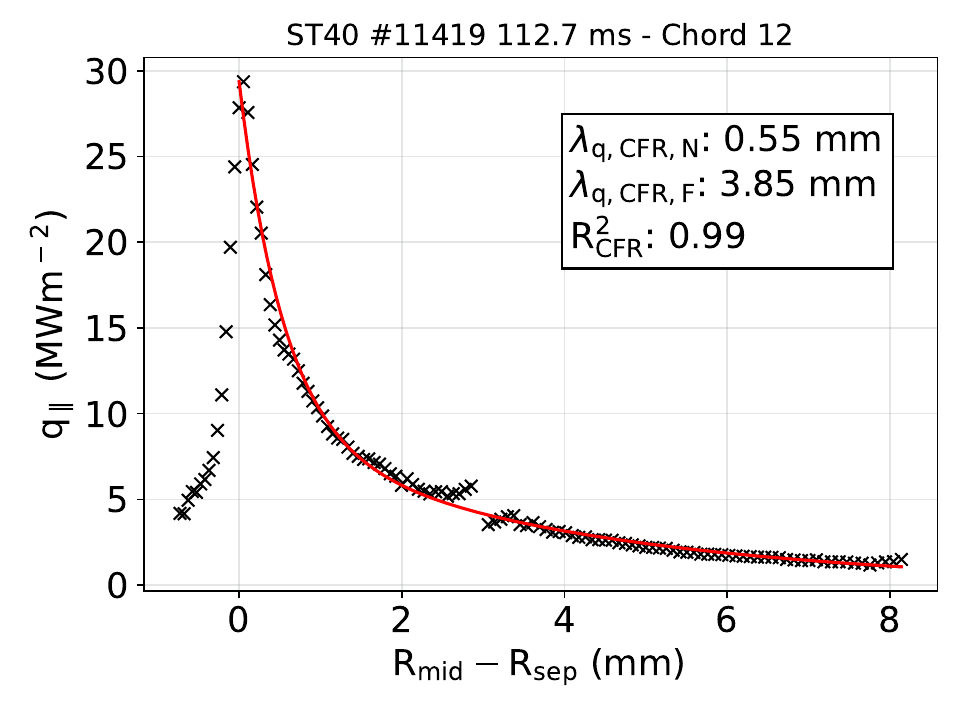}
    \caption{A double exponential profile is fitted to parallel heat flux data in the CFR for a single observation chord. Black crosses show the experimental data, with the red line representing the fit to the data. The resultant near and far e-folding lengths $\lambda_{q,N}$ and $\lambda_{q,F}$, along with the $R^{2}$ fit quality value are shown. These data lack error bars as the full propagation of errors throughout this IR analysis workflow is still under implementation.}
    \label{fig:ST40_double_fit}
\end{figure}

When taking a measure of the heat flux width of the SOL, it is common to refer to an e-folding length scale over which the parallel heat flux decays. Such a length scale, referred to as $\lambda_{q}$, can be extracted from profiles of parallel heat flux by fitting a function to these data. IRRITANT's flexibility allows for different functional forms to be fit to parallel heat flux data. Two such functions are an \enquote{Eich} profile\cite{eich2011inter}, which represents the heat flux as a radial exponential decay in the SOL that is convoluted with a Gaussian representing cross-field transport along the divertor leg, as well as a double exponential profile\cite{brunner2018high}, which represents the radial profile of heat flux in the CFR as a 2 exponential decay. In the case of the double exponential decay, 2 measures of the heat flux width are obtained, $\lambda_{q,N}$ representing the near SOL (close to the peak) and $\lambda_{q,F}$ representing the far SOL. An example of a double exponential fit along an observation chord is shown in figure \ref{fig:ST40_double_fit}.\newline

When fitting a function to the experimental data it is important to ensure that a suitable fit is achieved. To aid in this the LMFIT non-linear optimisation and curve fitting package is employed\cite{newville2016lmfit}. Due to the large number of fit parameters in these functions there is a risk in making use of non-global optimisation algorithms, as these can become stuck in local minima and exhibit greater sensitivity to initial value guesses. LMFIT provides convenient high-level access to various global optimisation algorithms which are better suited to probing a large parameter space and are less likely to become stuck in local minima. In this work we employ the \enquote*{basinhopping} global optimisation algorithm for performing curve fitting. This algorithm uses a random seed each time it is run to determine the points in parameter space that are selected for assessment. Because of this random seeding, the fit returned can vary when ran multiple times. As such, each time a profile of parallel heat flux for a given observation chord is to be fitted, a number of repeat fits are performed on that chord's data, with the $R^{2}$ fit quality value recorded for each repeat fit. It is found that the set of fit parameters across the repeat fittings will tend to cluster together, indicative of an optimal point in the space of the fit parameter. Any unsuccessful fits will return fit parameter values away from this cluster, which are in turn discarded based on their low $R^{2}$ value. Of those fit parameters that are left, a simple inter-quartile range filter is applied to remove any remaining outliers, before a mean value of the set of remaining parameter values is taken to be selected as the returned value of the fit parameter for the observation chord in question. A new $R^{2}$ value is then calculated for these final fit parameters.

As it is not known ahead of time what functional form will best represent the heat flux data, this entire fitting procedure is carried out twice, once using the Eich form and again separately with the double exponential form. In instances where the Eich form does not well represent the data being fit, all of the repeat fits being carried out on a given chord will yield low $R^{2}$ values. In these cases, the fit filtering process will yield parameters that lead to a poor fit to the data. The corresponding low $R^{2}$ is still recorded, and is later used to help identify times in which the heat flux profile on the divertor exhibits a narrow near-SOL feature, during which the $R^{2}$ from the double exponential fit will be high.\newline

With the fit parameters recorded for a given chord, a single set of fit parameters is also separately recorded that is intended to be indicative of the entire set of fit parameters across all of the observation chords. This allows for a single value of the heat flux width to be quoted for each time instance. The aforementioned fit filtering process is repeated on the set of fit parameters built from the fit parameter values returned by each chord individually. It is important to repeat this process across the set of chords so as to filter out any chords where, in spite of the above procedure, an acceptable fit was not achieved. This can occur due to the presence of hotspots on the divertor, arising from small indentations in the surface of the tiles, or on occasion from the curve fitting procedure failing to find a suitable minima, despite the repeat fittings performed.

\section{Heat flux distribution across the divertor}

The divertor surface heat flux $q_{\perp}$ on ST40 exhibits a low degree of axisymmetry for several reasons. Figure \ref{fig:ST40_divertor_chords} shows how the divertor strike band is slanted across the tile surfaces, this being the result of how the 3D geometry of the divertor intersects with the nominally axisymmetric flux surfaces of the plasma. Furthermore, this geometry results in the strike angle $\gamma$ varying across the tile surfaces. These two factors result in the magnitude of $q_{\perp}$ varying across the tiles for points of the same value of $\psi$, as well as the point of peak heat flux along the observation chords being at different poloidal locations.

\begin{figure}
    \centering
    \includegraphics[width = 0.95\linewidth]{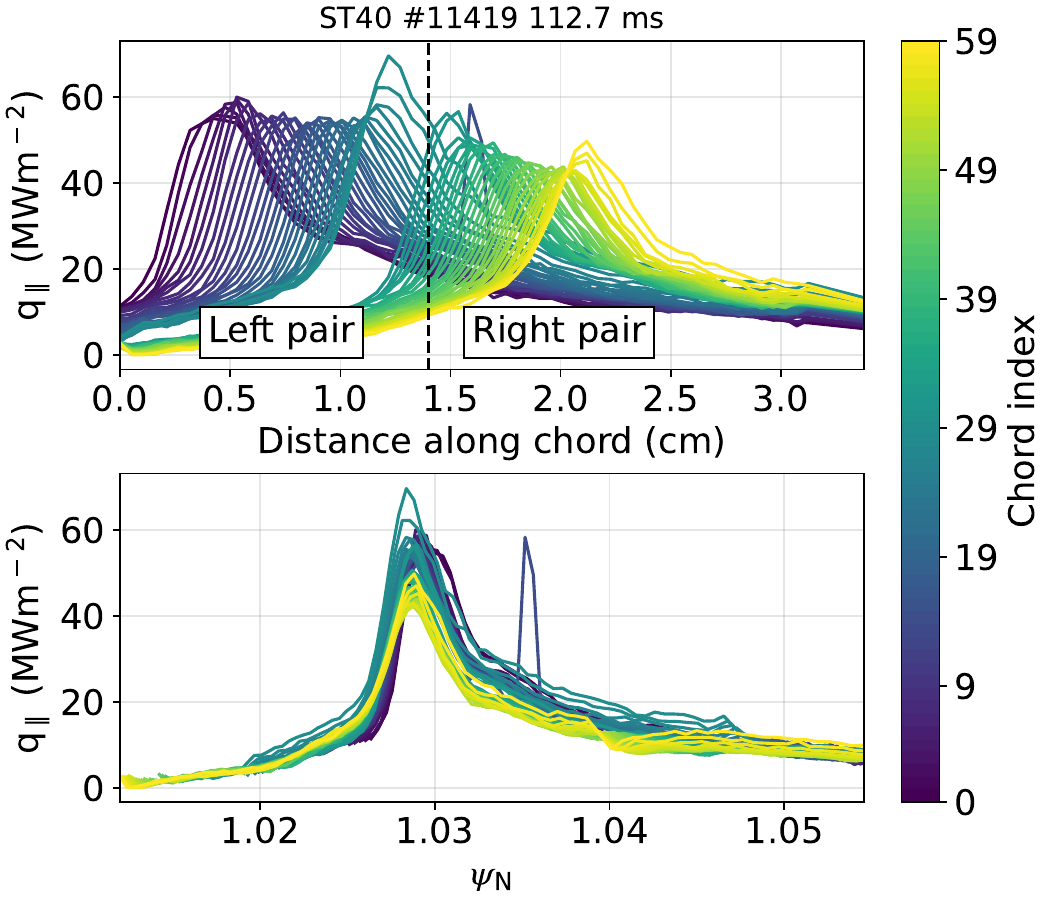}
    \caption{Parallel heat flux at the divertor target, $q_{\parallel,t}$ as a function of the distance along the divertor observation chord (top) and the normalised poloidal magnetic flux $\psi_{N}$ (bottom). The colour of each profile denotes the index of the chord, as illustrated in figure \ref{fig:ST40_divertor_chords}.}
    \label{fig:ST40_divertor_qpar}
\end{figure}

As shown in figure \ref{fig:ST40_divertor_qpar} the point of peak parallel heat flux is shifted along each observation chord, with this shift increasing with chord index. Also shown is the alignment of these peaks in $\psi$. For the left-hand tile pair (corresponding to chords 0-29), it is observed that for those chords not near a tile edge, there is an overlapping of the entire parallel heat flux profile in $\psi$. Interestingly, an uptick in the parallel heat flux along the entire profile is observed on those chords situated close to the trailing/leading edges of the left-hand tile pair. It is expected that on the poloidal chamfer adjacent to the leading edge there will be an increased surface heat flux (and hence temperature) due to the increase in $\gamma$ on the chamfered edge. Furthermore, one would expect for a given surface heat flux that the edges (both leading and trailing) of the tiles would exhibit an increased temperature pick up, due to the reduced thermal mass of the tile at its edges. At this stage it is not clear whether the observed increase in parallel heat flux at these edges is genuine, or is instead an artefact of the use of a 2D thermal that is not fully capturing heat transfer through the tile in a sufficiently accurate manner near the edges. Investigations into the accuracy of such a 2D model are underway and will be the subject of a future publication, where comparison against commercial software and 3D thermal models will be made. These observations for this left-hand tile pair suggest that the parallel heat flux arriving on the tile could be described by a function of $\psi$, if one puts aside the heat flux at the tile edges.\newline

\begin{figure}[hbt!]
    \centering
    \includegraphics[width = 0.95\linewidth]{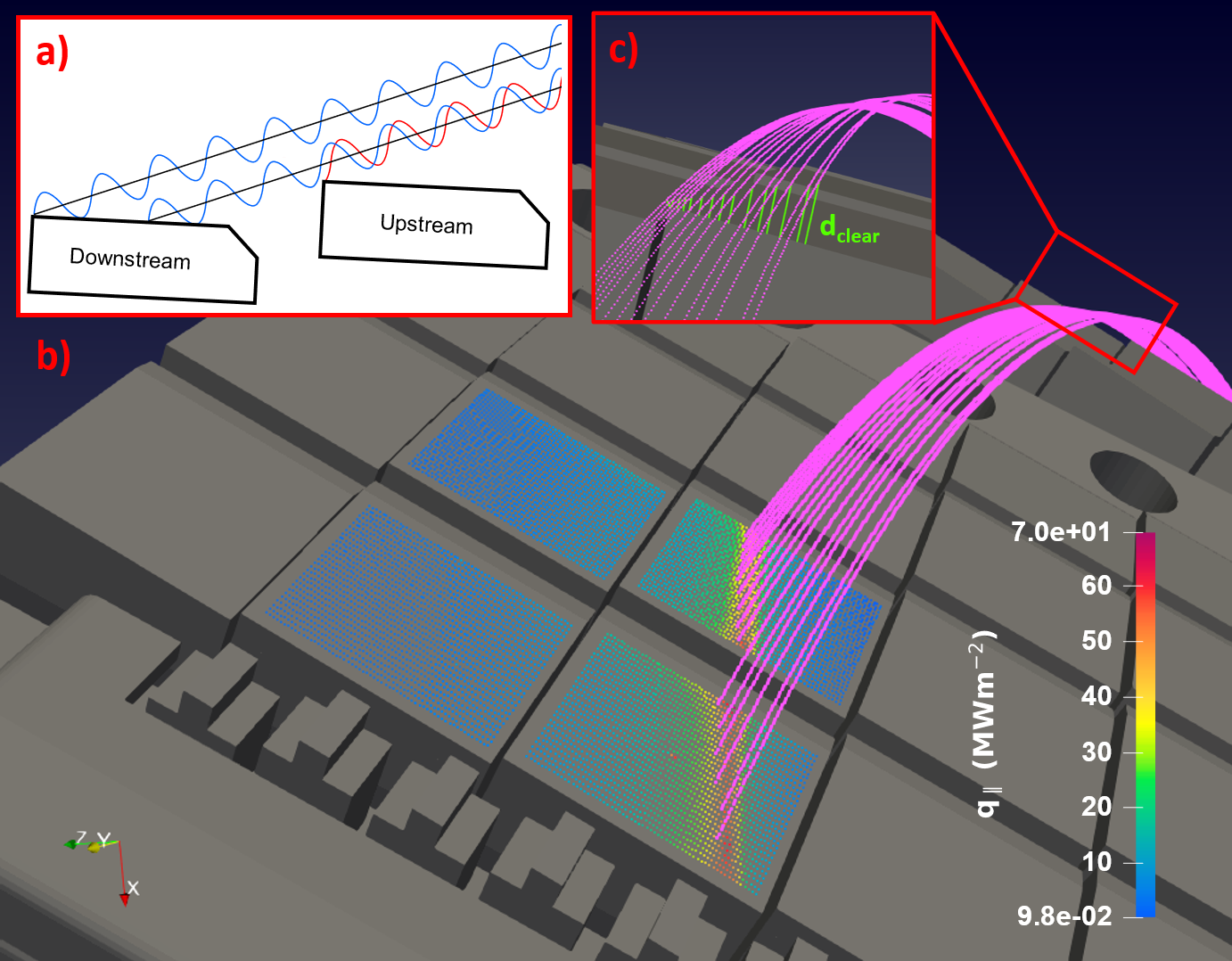}
    \caption{a) Illustration of partial shadowing on a pair of fish-scaled divertor tiles. b) Tracing of field lines (pink) across the strike band (coloured). c) Traced field lines passing over an upstream tile trailing edge. Field lines connected to the left tile pair (as viewed in figure \ref{fig:ST40_divertor_chords}) have a larger clearance distance (green) that those field lines connected to the right tile pair.}
    \label{fig:ST40_GYRO}
\end{figure}

The parallel heat flux pattern observed for the right-hand tile pair is similar to that of the left-hand tile pair, with an uptick in parallel heat flux at the leading/trailing edges. However, it is observed that as a whole the parallel heat flux is lower for the right-hand tile pair, with this heat flux dropping as one moves from the trailing edge of the tile towards the leading edge. One possible reason for this is to be attributed to the finite gyro-radius of ions as they precess along field lines whilst moving towards the divertor target. Figure \ref{fig:ST40_GYRO}a is an illustration of 2 divertor tiles in a fish-scaled configuration, with 2 field lines connected to the downstream target shown. Ions move from the upstream towards the target in a helical trajectory whose guiding centres are these magnetic field lines, gyrating about them at a distance referred to as the ions' Larmor radius, $r_{L}$. The left-most field line clears the trailing edge of the upstream tile at a distance greater than the ion Larmor radius, with these ions moving unimpeded from the upstream to the target. The right-most field line however clears the upstream trailing edge at a distance comparable to the Larmor radius. When this occurs the ions will either 'hop-over' the trailing edge, or, dependent upon their gyro-phase, they will instead strike the tile's trailing edge. As a result of this, a fraction of the ion flux is intercepted, leading to a reduction in heat flux from ions at the target when compared to those regions on the target near the left-most field line. This reduction in the target heat flux is referred to as partial shadowing.\newline

An investigation into whether partial shadowing could be responsible for the reduced parallel heat flux on the right-hand tile pair was carried out using the Heat flux Engineering Analysis Toolkit, HEAT\cite{looby2022software}. HEAT makes use of the field line tracing code MAFOT\cite{wingen2009high} which allowed for field lines across the strike band on the divertor at points of the same $\psi$ to be traced back upstream from the tile surface, as shown in figure \ref{fig:ST40_GYRO}b. The distance of closest approach between these field lines and the trailing edges of tiles situated on the next divertor segment upstream of the one being observed was calculated, as shown in figure \ref{fig:ST40_GYRO}c. As one moves from left to right across the observed tile pairs (increasing in chord index), these field lines have a reduced clearance distance $d_{clear}$ with the upstream tiles. For the left hand tile pair, this distance is typically $>$5mm, whereas for the right hand tile pair, as one moves from the trailing towards the leading edge, this clearance decreases. In order for partial shadowing to occur here, the ion Larmour radius would need to be at least equal to half of this clearance distance. This is so that an ion whose guiding centre is a field line situated midway between each traced field line and the upsteam tile's trailing edge can, depending on its gyro-phase, either hop-over or strike the edge, hence leading to partial shadowing at the downstream target. One can estimate the temperature of an ion with such a Larmor radius as a function of these clearance distances

\begin{equation}\label{eq:temp}
    T_{ion}(eV) = \frac{q_{ion}^{2}B^{2}r_{L}^{2}}{2m_{ion}e}
\end{equation}

where $q_{ion}$ is the charge of the ion, $m_{ion}$ is its mass, $r_{L}=\frac{d_{clear}}{2}$ and B is the strength of the magnetic field near the trailing edge, taken to be $\sim$1.85 T. For those field lines on the right hand tile pair, $d_{clear}$ was up to 2.5 mm, with a corresponding required ion temperature of up to 128 eV. Such a range of temperatures is conceivable in the divertor region of ST40, and as such suggests that the reduced parallel heat flux observed on the right-hand tile pair is due to partial shadowing that results from ions scraping off onto the upstream trailing edges of tiles. Future work in modelling the ST40 divertor will probe this further, and will involve the use of the gyro-orbit module within HEAT\cite{looby20223d}, which has the ability to trace ion macroparticles from an upstream source down to the divertor target along field lines, with ions scraping off onto PFCs upstream of the target location if their gyro-radius and phase allow for it.

\section{Conclusion}
A new IR thermography analysis toolchain has been implemented on ST40, made possible by the recent developments of diverted plasma configurations and the commissioning of a new high-resolution IR camera. Measurements of the surface heat flux on a section of the LFS region on the upper divertor are combined with information on the 3D shape of its surfaces and the magnetic topology of the plasma equilibrium, in order to calculate the heat flux parallel to magnetic field lines in the SOL region. These heat flux profiles are taken along several observation chords across the toroidal extent of the divertor, and are fit with functions of the form of both double and single exponential decays, in order to extract measurements of the width of the SOL. Measurements are aggregated across observation chords in order to infer a single value taken as the SOL width measurement. With the advent of this IR thermography workflow, measurements of the SOL heat flux width are now made possible in ST40. A decrease in the parallel heat flux across part of the divertor target region is observed, with this attributed to gyro-orbits scraping off on trailing edges upstream of this region, which will be further explored in follow up work. A statistical analysis of SOL width measurements in ST40 is explored in\cite{zhang2024st40}.

\section{Acknowledgements}
The ST40 divertor IR camera was procured and commissioned by collaborators from Oak Ridge National Laboratory as part of the CRADA NFE-19-07769 between Tokamak Energy and the U.S. Department of Energy. This submitted manuscript has been co-authored by a contractor of the U.S. Government under contract DE-AC05-00OR22725. Accordingly, the U.S. Government retains a nonexclusive, royalty-free license to publish or reproduce the published form of this contribution, or allow others to do so, for U.S. Government purposes.

\bibliography{biblio.bib}

\begin{thebibliography}{10}
\expandafter\ifx\csname url\endcsname\relax
  \def\url#1{\texttt{#1}}\fi
\expandafter\ifx\csname urlprefix\endcsname\relax\def\urlprefix{URL }\fi
\expandafter\ifx\csname href\endcsname\relax
  \def\href#1#2{#2} \def\path#1{#1}\fi

\bibitem{pitts2019physics}
R.~A. Pitts, X.~Bonnin, F.~Escourbiac, H.~Frerichs, J.~Gunn, T.~Hirai, A.~Kukushkin, E.~Kaveeva, M.~Miller, D.~Moulton, et~al., Physics basis for the first {ITER} tungsten divertor, Nuclear Materials and Energy 20 (2019) 100696.

\bibitem{kuang2020divertor}
A.~Kuang, S.~Ballinger, D.~Brunner, J.~Canik, A.~Creely, T.~Gray, M.~Greenwald, J.~Hughes, J.~Irby, B.~LaBombard, et~al., Divertor heat flux challenge and mitigation in {SPARC}, Journal of Plasma Physics 86~(5) (2020) 865860505.

\bibitem{vondracek2019divertor}
P.~Vondracek, E.~Gauthier, M.~Grof, M.~Hron, M.~Komm, R.~Panek, et~al., Divertor infrared thermography on {COMPASS}, Fusion Engineering and Design 146 (2019) 1003--1006.

\bibitem{adamek2017electron}
J.~Adamek, J.~Seidl, J.~Horacek, M.~Komm, T.~Eich, R.~Panek, J.~Cavalier, A.~Devitre, M.~Peterka, P.~Vondracek, et~al., Electron temperature and heat load measurements in the {COMPASS} divertor using the new system of probes, Nuclear Fusion 57~(11) (2017) 116017.

\bibitem{hecko2023experimental}
J.~Hecko, M.~Komm, M.~Sos, J.~Adamek, P.~Bilkova, K.~Bogar, P.~Bohm, F.~Jaulmes, I.~Mysiura, M.~Tomes, et~al., Experimental evidence of very short power decay lengths in {H}-mode discharges in the {COMPASS} tokamak, Plasma Physics and Controlled Fusion 66~(1) (2023) 015013.

\bibitem{mcnamara2024st40}
S.~McNamara, et~al., Overview of recent results from the {ST40} compact high-field spherical tokamak, Submitted to Nuclear Fusion (2024).

\bibitem{bamber2021st40}
R.~Bamber, D.~Iglesias, O.~Asunta, P.~Bunting, S.~Daughtry, G.~Dunbar, S.~Hanks, A.~Horozaniecki, P.~Moore, D.~Lockley, et~al., The {ST40} {IVC1} divertor project: Procurement and installation in times of {COVID-19}, Fusion Engineering and Design 168 (2021) 112378.

\bibitem{calcam}
S.~Silburn, J.~Harrison, T.~Farley, J.~Cavalier, S.~V. Stroud, J.~McGowan, A.~Marignier, E.~Nurse, C.~Gutschow, M.~Smithies, A.~Wynn, R.~Doyle, M.~Kriete, A.~Perek, \href{https://zenodo.org/doi/10.5281/zenodo.10655746}{Calcam} (2024).
\newblock \href {https://doi.org/10.5281/ZENODO.10655746} {\path{doi:10.5281/ZENODO.10655746}}.
\newline\urlprefix\url{https://zenodo.org/doi/10.5281/zenodo.10655746}

\bibitem{robinson2024fahf}
M.~Robinson, M.~Maartensson, M.~Moscheni, A.~Rengle, M.~Jackson, V.~Lee, E.~Yildirim, T.~Gray, Poster: {ST40} tool for infrared thermography: {FAHF}, in: 26th International Conference on Plasma Surface Interaction, 2024.

\bibitem{lao1985reconstruction}
L.~Lao, H.~S. John, R.~Stambaugh, A.~Kellman, W.~Pfeiffer, Reconstruction of current profile parameters and plasma shapes in tokamaks, Nuclear fusion 25~(11) (1985) 1611.

\bibitem{FreeCAD}
\href{https://www.freecad.org/}{[link]}.
\newline\urlprefix\url{https://www.freecad.org/}

\bibitem{Perronnet}
A.~Perronnet, \href{https://www.ljll.fr/perronnet/mefistoa.gene.html}{Mefisto}.
\newline\urlprefix\url{https://www.ljll.fr/perronnet/mefistoa.gene.html}

\bibitem{eich2011inter}
T.~Eich, B.~Sieglin, A.~Scarabosio, W.~Fundamenski, R.~J. Goldston, A.~Herrmann, A.~U. Team, et~al., {Inter-ELM} power decay length for {JET} and {ASDEX Upgrade}: measurement and comparison with heuristic drift-based model, Physical review letters 107~(21) (2011) 215001.

\bibitem{brunner2018high}
D.~Brunner, B.~LaBombard, A.~Kuang, J.~Terry, High-resolution heat flux width measurements at reactor-level magnetic fields and observation of a unified width scaling across confinement regimes in the {A}lcator {C}-{M}od tokamak, Nuclear Fusion 58~(9) (2018) 094002.

\bibitem{newville2016lmfit}
M.~Newville, T.~Stensitzki, D.~B. Allen, M.~Rawlik, A.~Ingargiola, A.~Nelson, {LMFIT}: Non-linear least-square minimization and curve-fitting for {P}ython, Astrophysics Source Code Library (2016) ascl--1606.

\bibitem{looby2022software}
T.~Looby, M.~Reinke, A.~Wingen, J.~Menard, S.~Gerhardt, T.~Gray, D.~Donovan, E.~Unterberg, J.~Klabacha, M.~Messineo, A software package for plasma-facing component analysis and design: The heat flux engineering analysis toolkit ({HEAT}), Fusion Science and Technology 78~(1) (2022) 10--27.

\bibitem{wingen2009high}
A.~Wingen, T.~Evans, K.~Spatschek, High resolution numerical studies of separatrix splitting due to non-axisymmetric perturbation in {DIII-D}, Nuclear fusion 49~(5) (2009) 055027.

\bibitem{looby20223d}
T.~Looby, M.~Reinke, A.~Wingen, T.~Gray, E.~Unterberg, D.~Donovan, {3D} ion gyro-orbit heat load predictions for {NSTX-U}, Nuclear Fusion 62~(10) (2022) 106020.

\bibitem{zhang2024st40}
X.~Zhang, et~al., Experimental evidence of bifurcated power decay lengths in the near scrape-off layer of tokamak plasma, Under preparation (2024).

\end{thebibliography}

\end{document}